\documentclass[preprint,aps,12pt,preprintnumbers,eqsecnum,nofootinbib]{revtex4}
\usepackage{graphicx}
\usepackage{subfigure}

\usepackage{color}
\usepackage{amssymb,amsmath}
\def\lag{{\mathcal{L}}}

\newcommand{\beq}{\begin{equation}}
\newcommand{\enq}{\end{equation}}

\begin{document}
%
%
\title{\vspace*{0.5in} 
Classical scale-invariance, the electroweak scale \\ and vector dark matter
\vskip 0.1in}
\author{Christopher D. Carone}\email[]{cdcaro@wm.edu}
\author{Raymundo Ramos}\email[]{raramos@email.wm.edu}

\affiliation{High Energy Theory Group, Department of Physics,
College of William and Mary, Williamsburg, VA 23187-8795}
\date{July 2013}

\begin{abstract}
We consider a classically scale-invariant extension of the standard model in which a dark, non-Abelian gauge 
symmetry is spontaneously broken via the Coleman-Weinberg mechanism. Higgs portal couplings between the
dark and standard model sectors provide an origin for the Higgs mass squared parameter and, hence, the
electroweak scale.  We find that choices for model parameters exist in which the dark gauge multiplet is viable as 
dark matter.  
\end{abstract}
\maketitle
\section{Introduction} \label{sec:intro}
Over the past decade, solutions to the hierarchy problem have been dominated by an appealing
theoretical paradigm: partners to standard model particles are postulated to cancel the 
quadratic divergence that otherwise affects the Higgs boson squared mass.  These partners can
have spins that differ from those of their standard model counterparts, as in the
minimal supersymmetric standard model (MSSM)~\cite{mssm}, or the same spins, as in little higgs 
models~\cite{lhm}.  They can be associated with states in Hilbert space of positive norm, as in the preceding
two examples, or states of negative norm, as in the Lee-Wick standard model~\cite{lwsm}.  A point of commonality
in all these scenarios is the requirement that the partner particles appear at or near the electroweak
scale, which one might reasonably identify with the Higgs field vacuum expectation value (vev), 
$v=246$~GeV.  Searches at the Large Hadron Collider (LHC) for new particles around this energy scale have, aside
from the Higgs boson, produced null results~\cite{null}.

Of course, all the scenarios described in the preceding paragraph have a decoupling limit, and it is
a matter of taste how much fine-tuning one is willing to tolerate before concluding that a given proposal is 
disfavored.  One might hope that the planned energy upgrade at the LHC will provide more definitive 
results.  Nevertheless, the absence of even small indirect effects of partner particles in the current 
LHC data motivates the study of alternative paradigms. Here, we consider a scenario first discussed by 
Bardeen~\cite{bardeen}, and studied recently by many others~\cite{older,old,recent,strumx,dsusy}, that the standard 
model may possess a softly-broken classical scale-invariance that protects it from unwanted quadratic
divergences. Such a scenario can be realized if the standard model Lagrangian has no dimensionful parameters 
and the Higgs mass arises via dimensional transmutation.  This can occur if the Higgs field couples to a new strongly 
interacting sector, as explored in Refs.~\cite{darkstrong}.  (For a much earlier example of a classically scale-invariant 
theory in which the Higgs boson mass is determined via dimensional transmutation in a strongly interacting sector, 
see Ref.~\cite{cg}.)  Alternatively, the Higgs boson mass can arise in a classically scale-invariant theory that is weakly 
coupled via the Coleman-Weinberg (CW) mechanism~\cite{cw}.  It is well known that the CW mechanism applied to the 
standard model alone leads to a Higgs boson mass that is much smaller than the electroweak gauge boson masses, and 
hence is not viable.  However, Refs.~\cite{older,old,recent} demonstrate explicitly that modest extensions of the standard 
model can avoid this problem. It is this general approach that we pursue in the model building discussed in this paper.

The argument of Bardeen has been rephrased a number of times in Refs.~\cite{old,recent}, with additional 
justification and emphasis varying from paper to paper (See also a summary given in a talk by Lykken~\cite{lykken}).  
Rather than repeating this discussion, we refer the reader to these references; here make only a few comments.   In order 
for an extension of the standard model to be classically scale invariant and free of quadratic divergences, one first 
assumes that the tree-level Higgs mass term is absent and  that there are no higher mass scales associated with new heavy 
particle thresholds, as would be the case, for example, in a grand unified theory.  The latter requirement precludes a conventional 
see-saw mechanism for the generation of small neutrino masses, so we will simply assume that neutrinos have Dirac 
mass terms with small Yukawa couplings.  As in the charged fermion sector, small neutrino masses are then 
technically natural~\cite{thooft} since chiral symmetries are restored in the limit of vanishing Yukawa couplings.  As flavor 
physics is not the focus of the present work, this assumption will suffice for the present purposes.  If one then 
works with a regulator that does the least violence to the classical symmetry (namely, dimensional regularization), then one
observes that a Higgs mass squared generated radiatively in the infrared is only multiplicatively renormalized~\cite{recent};  this indicates that it too is technically natural.  The only remaining assumption is that quantum 
gravitational physics does not spoil this outcome even though it is associated with a dimensionful scale, {\em viz.}, 
the Planck scale $M_{Pl}=1.22 \times 10^{19}$~GeV (or alternatively, the reduced Planck scale, $M_*=2.43 
\times 10^{18}$~GeV).  Our current uncertainty about the nature of quantum gravity makes this at most a 
plausible working assumption, but one that leads to a relatively restrictive framework for low-energy model 
building.  Such models can be more readily put to direct experimental tests.

The model we study is one in which the standard model is extended by an additional  SU(2)$_D$ 
gauge group and a complex scalar doublet $\Phi$ that transforms only under this new gauge symmetry.   The 
subscript represents the word ``dark", since the new gauge sector only communicates with the standard model 
via a coupling between $\Phi$ and the standard model Higgs doublet field $H$, 
\begin{equation}
\lambda_p \,\Phi^\dagger \Phi H^\dagger H \,\,\, ,
\label{eq:portal1}
\end{equation}
where $\lambda_p$ is typically small.  This is the well-known Higgs portal, one of the small number of possible 
renormalizable couplings between standard model fields and a new sector of particles that are singlets under
the standard model gauge group.  In the present case, the dark sector is scale invariant at tree level and 
undergoes spontaneous symmetry breaking via the Coleman-Weinberg mechanism.  Hence, the vev 
$\langle \Phi \rangle \equiv v_D/\sqrt{2}$, which provides an origin for the Higgs boson mass scale via 
Eq.~(\ref{eq:portal1}), is determined by dimensional transmutation.   The SU(2)$_D$ gauge group is
spontaneously broken, leading to a degenerate triplet of massive gauge bosons $A^a$, $a=1 \ldots 3$, with 
masses $m_A \equiv g_D v_D/2$, where $g_D$ is the SU(2)$_D$ gauge coupling.  These spin-one states are 
stable and are potential dark matter candidates.  One of the results we present in this paper is that there are 
parameter choices consistent with viable Coleman-Weinberg symmetry breaking as well as the correct relic density 
of the SU(2)$_D$ gauge multiplet.

The motivation for the work we present on this model can also be framed in the context of the existing related 
literature.  Let us give three different rationales that may appeal to readers with different theoretical tastes:

{\em i}.)  The use of the Higgs portal as a means for communicating Coleman-Weinberg symmetry breaking
in a dark sector to the standard model has been discussed recently in the context of Abelian dark gauge groups in 
Refs.~\cite{recent}.  Our work considers the phenomenology in a model based on a non-Abelian dark gauge 
group, a natural alternative possibility.

{\em ii}.)  The possibility that dark matter may be spin-one is well known, and the case in which the dark matter is a 
massive SU(2) gauge multiplet has been considered in Refs.~\cite{vdm}.  In this scenario, called Hidden Vector Dark Matter, 
the doublet field $\Phi$ together with the $H$ are assumed to have the most general scalar potential.  Our work studies the 
Coleman-Weinberg limit of the potential, leading to a model that is parametrically simpler and whose phenomenology is 
more constrained.

{\em iii}.)  There has been interest in dark matter models in which the dark matter candidate can annihilate
predominantly into lighter, unstable intermediate particles.  These ``secluded dark matter" scenarios~\cite{secluded} 
are less constrained by direct dark matter searches, since the annihilation cross section and the dark matter-nucleon 
elastic scattering cross section are determined by different combinations of couplings.  Our work studies a simple model 
that falls into this interesting category. 

Our paper is organized as follows. In Sec.~\ref{sec:model}, we define the model and our conventions.
In Sec.~\ref{sec:pheno}, we consider phenomenological constraints on the model, including
vacuum stability, perturbativity, and some aspects of Higgs boson physics.  In Sec.~\ref{sec:vdm}, we 
consider the parameter ranges in which the model can provide a viable vector dark matter candidate.  In
Sec.~\ref{sec:conc}, we summarize our conclusions.

\section{The Model} \label{sec:model}
The gauge symmetry of the model is $G_{\rm SM} \times $SU(2)$_D$, where $G_{\rm SM}$ is the standard
model gauge group.  The standard model particle content is assumed to include three right-handed neutrinos so 
that neutrino Dirac masses are possible, for the reasons described in the introduction.  In addition, the model
includes a complex scalar doublet under SU(2)$_D$.  No fermions transforming under
the dark gauge group are present, so the model is free of gauge anomalies.  

At tree-level, the scalar potential is given by
\begin{equation}
V(\Phi,H) = \frac{1}{2} \lambda\, (\Phi^\dagger \Phi)^2 - \lambda_p\, (H^\dagger H) (\Phi^\dagger \Phi)
+\frac{1}{2} \lambda_H\, (H^\dagger H)^2 \,\,\, ,
\label{eq:treepot}
\end{equation}
where $H$ is the standard model Higgs doublet field.  Mass terms for the $\Phi$ and $H$ fields
are omitted, in accordance with the assumption of classical scale invariance. Note that Eq.~(\ref{eq:treepot}) can be 
rewritten
\begin{equation}
V(\Phi,H) = \frac{1}{2} \lambda_H \left( H^\dagger H - \frac{\lambda_p}{\lambda_H} \Phi^\dagger \Phi\right)^2
+\frac{1}{2} \left(\lambda - \frac{\lambda_p^2}{\lambda_H}\right) (\Phi^\dagger \Phi)^2 \,\,\, ,
\label{eq:treepot2}
\end{equation}
from which one can read off the tree-level vacuum stability conditions
\begin{equation}
\lambda_H > 0  \,\,\,\,\,\,\,\,\,\, \mbox{ and } \,\,\,\,\,\,\,\,\, \lambda \, \lambda_H > \lambda_p^2 \,\,\, .
\label{eq:treevs}
\end{equation}
We will refer to these conditions again later in our analysis.

Given the absence of dimensionful couplings, it is not surprising that minimization of Eq.~(\ref{eq:treepot}) gives 
$\langle \Phi \rangle=\langle H \rangle=0$.   This outcome, however, does not persist when quantum 
corrections to $V(\Phi,H)$ are taken into account~\cite{cw}.  We include the one-loop contributions to the effective
potential that involve the SU(2)$_D$ gauge bosons and the top quark.   For the numerical values of the couplings 
that are relevant in our later analysis, these represent the leading corrections.  Defining the classical fields
$\phi$ and $\sigma$ by 
\begin{equation}
\Phi = \frac{1}{\sqrt{2}}  \left(\begin{array}{c} 0 \\ \phi \end{array}\right) \,\,\,\,\,\,\,\,\,\,
\mbox{ and } \,\,\,\,\,\,\,\,\,\, H = \frac{1}{\sqrt{2}} \left(\begin{array}{c} 0 \\  \sigma \end{array}\right) \,\,\, ,
\end{equation}
the one-loop effective potential may be written
\begin{eqnarray}
V(\phi, \sigma)^{\overline{{\rm MS}}} &=& \frac{1}{8} \lambda\, \phi^4+\frac{9}{1024\pi^2} g_D^4 \phi^4
\left(\ln\frac{g_D^2 \phi^2}{4 \mu^2} - \frac{3}{2}\right)-\frac{1}{4} \lambda_p \, \sigma^2 \phi^2 \nonumber \\
&+& \frac{1}{8} \lambda_H \, \sigma^4 -\frac{3}{64\pi^2} h_t^4 \sigma^4 \left(\ln\frac{h_t^2 \sigma^2}{2 \mu^2} -\frac{3}{2}\right) \,
\,\, ,
\label{eq:epot1}
\end{eqnarray}
where $h_t$ is the top quark Yukawa coupling.   In Eq.~(\ref{eq:epot1}) we work in the $\overline{{\rm MS}}$
scheme and $\mu$ is the renormalization scale.   We extremize the potential by evaluating
\begin{equation}
\frac{\partial V}{\partial \phi} = \frac{\partial V}{\partial \sigma} =0 \,\,\, ,
\end{equation}
with $V$ and the couplings contained therein evaluated at the renormalization scale $\mu = \langle \sigma \rangle\equiv v$.  
(Note that we use this potential only to relate couplings defined at the electroweak scale and vevs that do not differ wildly
from the same scale. For this purpose, renormalization group improvement is not necessary to achieve reliable results.)
This leads to two constraints on the solution with nonvanishing $\langle \phi \rangle$ and $\langle \sigma \rangle$,
\begin{equation}
\lambda_H =  \lambda_p \frac{\langle \phi \rangle^2}{\langle \sigma \rangle^2} 
- \frac{3}{8 \pi^2} h_t^4 \left[1-\ln\left(\frac{h_t^2}{2}\right)\right] \,\,\, ,
\label{eq:dv1}
\end{equation}
\begin{equation}
\lambda = \frac{9}{128 \pi^2} g_D^4 \left[1-\ln\left(\frac{g_D^2 \langle \phi \rangle^2}{4 \langle \sigma \rangle^2}\right)\right]+ \lambda_p \frac{\langle \sigma \rangle^2}{\langle \phi \rangle^2} \,\,\, .
\label{eq:dv2}
\end{equation}
We fix $\langle \sigma \rangle \equiv v =246$~GeV, as indicated earlier, while $h_t = \sqrt{2} m_t / v $ follows numerically from the $\overline{{\rm MS}}$ value of the top quark mass, $m_t=160^{+5}_{-4}$~GeV~\cite{rpp}.  Thus far, Eqs.~(\ref{eq:dv1}) and 
(\ref{eq:dv2}) imply that one can take the free parameters of the model to be $g_D$, $\lambda_p$ and $\langle \phi \rangle$.  

We know, however, that one parametric degree of freedom is fixed by the requirement that one of the two scalar mass eigenstates must correspond to the Higgs boson observed at the LHC.  To proceed, we consider the scalar mass
squared matrix that follows from Eqs.~(\ref{eq:epot1}), (\ref{eq:dv1}) and (\ref{eq:dv2}):
\begin{equation}
M^2 = \left[ \begin{array}{cc}  \langle \phi \rangle^2 \lambda_p + \Delta m^2 & -\lambda_p \langle \sigma \rangle \langle \phi \rangle \\
-\lambda_p \langle \sigma \rangle \langle \phi \rangle & \frac{9}{128 \pi^2} g_D^4 \langle \phi \rangle^2 +\lambda_p \langle \sigma \rangle^2
\end{array} \right]  \,\,\, .
\label{eq:msmatrix}
\end{equation}
Here, $\Delta m^2 = -3 h_t^4 \langle \sigma \rangle^2 / (8 \pi^2)$ is the shift in the Higgs boson mass in the
standard model due to the top quark loop correction.  For given input values of $(g_D,\lambda_p)$, we 
solve for $\langle \phi \rangle$ numerically by identifying either eigenvalue of Eq.~(\ref{eq:msmatrix}) 
with the Higgs boson mass $m_h=125$~GeV.   The two choices correspond to either $m_\eta>m_h$ or 
$m_\eta<m_h$, where we let $\eta$ represent the other scalar mass eigenstate.
We define the mixing angle $\theta$ by
\begin{equation}
\left(\begin{array}{cc} \cos\theta & -\sin\theta \\ \sin\theta & \cos\theta \end{array}\right)
\left(\begin{array}{c} h \\ \eta \end{array}\right) = \left(\begin{array}{c} \sigma_0 \\ \varphi \end{array}\right) \,\,\, ,
\label{eq:mxmtx}
\end{equation}
where $\varphi$ and $\sigma_0$ are the physical fluctuations about the vevs: 
$\phi = \langle \phi \rangle+\varphi$ and $\sigma=\langle \sigma \rangle+\sigma_0$. It follows that 
$\tan 2\theta = 2 M^2_{12} / (M^2_{11}-M^2_{22})$, where $M^2_{ij}$ are the elements of the matrix in
Eq.~(\ref{eq:msmatrix}).

The phenomenology of the model may now be specified in terms of a two-dimensional parameter space,
the $(g_D, \lambda_p)$ plane.  We begin isolating interesting regions of this parameter space in the next
section.

\section{Phenomenological Constraints} \label{sec:pheno}
Given our assumption that there are no new, physical mass scales between the weak and Planck scales, we first 
require that viable points in parameter space do not lead to Landau poles below $M_*$ in any of the couplings.   This 
precludes the possibility that a Landau pole is a symptom of omitted new physics that is associated with an 
intermediate mass scale.  Of course, before a Landau pole is reached, a given coupling will become nonperturbatively 
large, and one cannot be sure that it actually blows up.  We simply impose the requirement that $\lambda$, 
$\lambda_H$ and $\lambda_p$ remain each smaller than $3$ below $\mu=M_*$.  We find that the allowed parameter
space of the model is not significantly enlarged for larger choices of this numerical limit, since couplings that exceed it
tend to do so very quickly.  To proceed, we numerically evaluate the following one-loop renormalization group equations (RGEs),
\begin{eqnarray}
&& 16 \pi^2 \frac{d \lambda}{dt}  =  12 \lambda^2 + 4 \lambda_p^2 - 9 g_D^2 \lambda+\frac{9}{4} g_D^4 \,\,\, ,
\end{eqnarray}
\begin{eqnarray}
16 \pi^2 \frac{d \lambda_H}{dt} &=&  12 \lambda_H^2 + 4 \lambda_p^2 
+\frac{9}{4} \left(\frac{3}{25} g_1^4 + \frac{2}{5} g_1^2 g_2^2 +g_2^4\right) \nonumber \\
&-&\left(\frac{9}{5} g_1^2+9 g_2^2\right) \lambda_H + 12 h_t^2 \lambda_H -12 h_t^4 \,\,\, ,
\end{eqnarray}
\begin{eqnarray}
&& 16 \pi^2 \frac{d \lambda_p}{dt}  =  \lambda_p \left(6 \lambda -4 \lambda_p+6 \lambda_H+6 h_t^2
-\frac{9}{2} g_2^2 - \frac{9}{10} g_1^2 - \frac{9}{2} g_D^2\right) \,\,\, ,
\end{eqnarray}
\begin{equation}
16 \pi^2 \frac{d g_D}{dt} = -\frac{43}{6}  g_D^3 \,\,\, .
\end{equation}
Here $h_t$, $g_1$, $g_2$ and $g_3$ are evolved according to the one-loop standard model RGEs
\begin{eqnarray}
&& 16 \pi^2 \frac{d h_t}{dt}  = \left[-\frac{17}{20} g_1^2 -\frac{9}{4} g_2^2 - 8 g_3^2 + \frac{9}{2} h_t^2\right] h_t  \,\,\, ,
\end{eqnarray}
\begin{equation}
16 \pi^2 \frac{d g_i}{dt} = b_i \,g_i^3 \,\,\, ,
\end{equation}
where $b_i = (\frac{41}{10},-\frac{19}{6}, -7)$ and we use the SU(5) normalization of hypercharge.  In addition to the
assignment of initial conditions described in Sec.~\ref{sec:model}, we take $\alpha_1^{-1}(m_Z)=59.02$, 
$\alpha_2^{-1}(m_Z)=29.58$ and $\alpha_3^{-1}(m_Z)=8.36$~\cite{rpp}.  Defining the parameter $t=\ln(\mu/m_Z)$, we evaluate
the RGE's between $t=0$ and $t_*=\ln(M_*/m_Z) \approx 37.8$, ignoring threshold corrections at the weak scale.
We note that our requirement that the couplings remain bounded everywhere in this interval may be overly 
conservative, since (as theories of TeV-scale gravity have illustrated) the cut off at which gravitational physics becomes 
relevant may in fact be substantially smaller than $M_*$.

\begin{figure}[b]
\subfigure{\includegraphics[width = 0.495\textwidth]{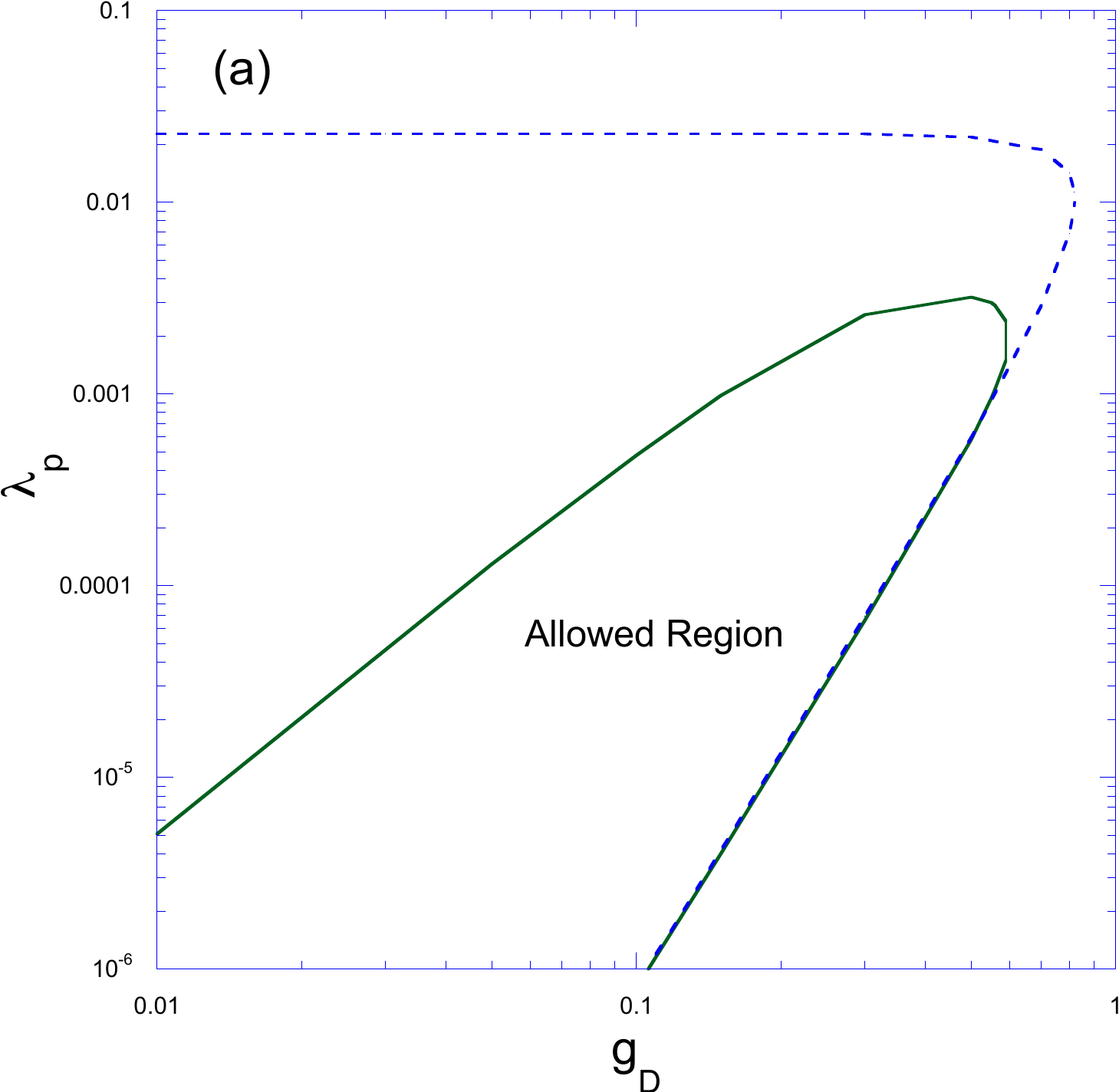}}
\subfigure{\includegraphics[width = 0.495\textwidth]{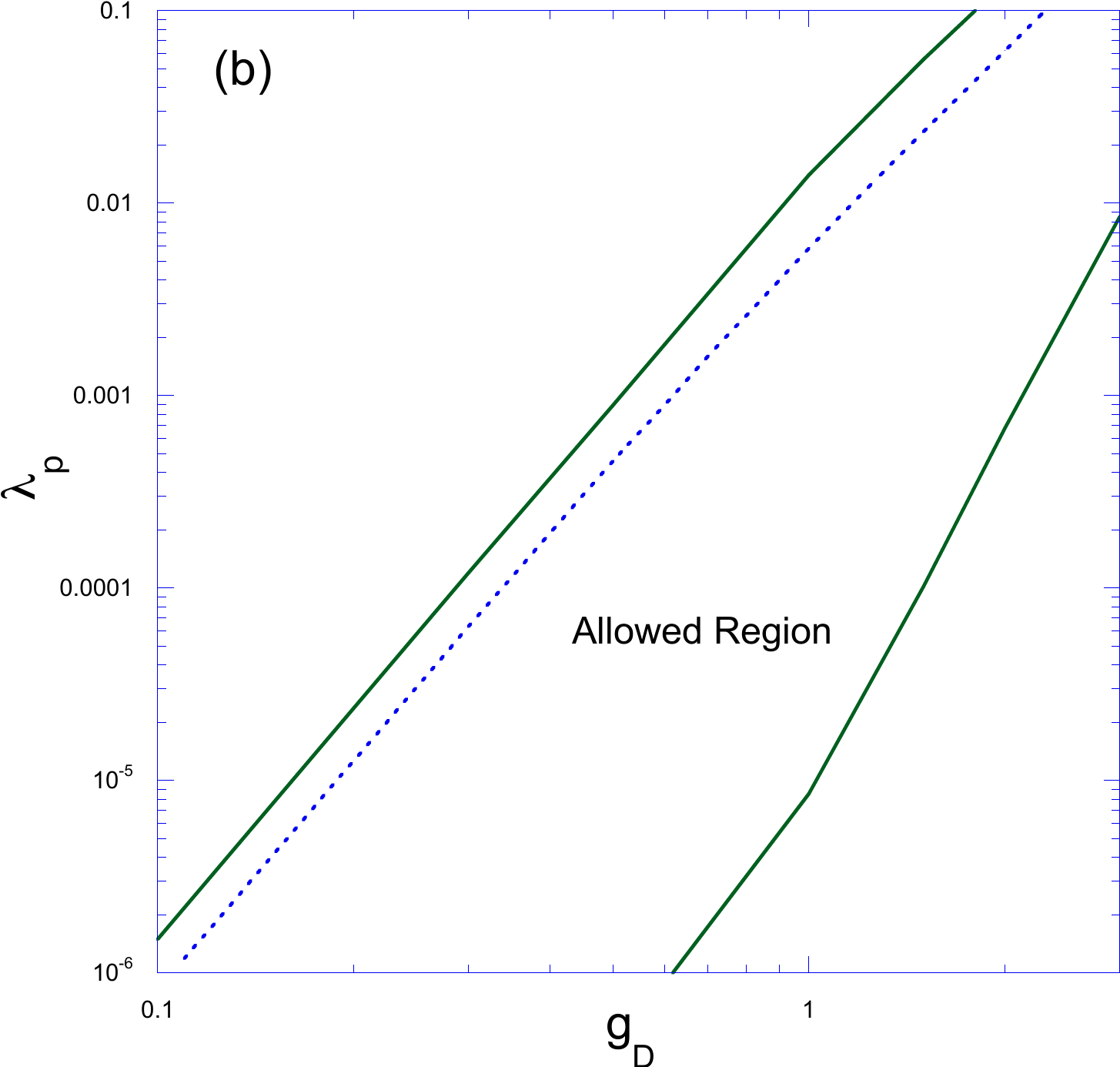}}
\caption{\label{fig:vsp} Regions of the $g_D$-$\lambda_p$ plane that are consistent with the perturbativity and
vacuum stability constraints discussed in the text. In (a), $m_\eta < m_h$, while in (b), $m_\eta > m_h$. The regions 
above and to the right of the dashed line in (a) and above the dashed line in (b) correspond to $\sin^2\theta>0.1$.}
\end{figure}
We are also now equipped to determine the vacuum stability of the model at each point in parameter space.  In the 
standard model, one runs the Higgs quartic coupling to higher renormalization scales and determines whether there
are points where the coupling becomes negative.  This result implies that the effective potential becomes
unbounded from below.  In two-Higgs doublet models, the standard approach is also to run the couplings of the 
tree-level potential, and to check that the tree-level stability conditions remain satisfied.  The justification for this procedure 
is discussed in some detail in Ref.~\cite{sher}.  Applying this approach to the present model, we require at large renormalization 
scales that Eq.~(\ref{eq:treevs}) remain satisfied.  As discussed in Ref.~\cite{krey}, we do not expect these conditions to be 
satisfied at small scales, since we know that at small $t$ the tree-level potential is {\em not} stable;  the one-loop 
corrections are necessary ingredients for obtaining vacuum stability in this region.  Given a choice of the two free
parameters, the values and signs of all the remaining couplings are determined.  Hence, our scan over parameter
space will include all possible values of the electroweak-scale couplings that are phenomenologically viable.  We then 
require that Eq.~(\ref{eq:treevs}) remain satisfied over some range $t_0 < t < t_*$ with $t_0$ sufficiently 
larger than zero to eliminate cases in which the potential turns over and becomes unbounded from below at large field values.  For 
definiteness, we take $t_0=5$ in computing our numerical results; our conclusions are not sensitive to the precise value of $t_0$.  
The allowed regions that remain after the constraints of perturbativity and vacuum stability are 
imposed are shown in Fig.~\ref{fig:vsp}. There is no simple qualitative explanation for the shapes of these regions.  
Eqs.~(\ref{eq:dv1}) and (\ref{eq:dv2}) as well as the RGEs are nonlinear; at some points in parameter space, 
$\lambda_H$ reaches a Landau pole before $M_*$, while at nearby points one of the other couplings is first to become 
unacceptably large or leads to a violation of a stability condition.  Note that we do not extend these plots to smaller values of 
$g_D$, since we will find that relatively large values of the couplings are required to obtain the desired dark vector 
annihilation cross section.  This will be discussed in the next section.

The remaining issue we wish to address in this section is Higgs boson physics.  The fact that the scalar mass eigenstates 
are mixtures of $\sigma_0$ and $\varphi$, where $\sigma_0$ would otherwise correspond to the standard model Higgs 
field, suggests that the model could lead to observable deviations of Higgs properties away from their standard model
expectations.  The production cross section times branching fractions of the Higgs-like eigenstate are proportional 
to $\cos^2\theta$ times their standard model values.  Current LHC bounds imply that this proportionality factor 
can be no smaller than $\approx 0.7$~\cite{ihb}.  Moreover, if the mixing is large, then the otherwise ``dark" 
Higgs $\eta$ would develop large enough couplings to the visible sector to be detected in Higgs boson searches at 
the LHC\footnote{For an interesting exception to this statement, see Ref.~\cite{twins}}. 
In this case, the partial widths to standard model quarks, leptons and gauge bosons are $\sin^2\theta$ 
times the value for a standard model Higgs.  Ignoring possible decay to two standard model Higgs, one would expect 
that the branching fractions for the $\eta$ state to be the same as a standard model Higgs boson, but the production 
cross section suppressed by a factor of $\sin^2\theta$.  LHC heavy Higgs search bounds can all be evaded for 
$\sin^2\theta \alt 0.1$~\cite{hhb}. Hence, we show in Fig.~\ref{fig:vsp} the regions in which $\sin^2\theta$ exceeds this value.  The true constraint is actually weaker (since the LHC bound is not as restrictive as $0.1$ for all scalar boson masses) but the distinction is not important here since the difference this produces in the allowed parameter region of Fig.~\ref{fig:vsp} is relatively small.

\section{Vector Dark Matter} \label{sec:vdm}
\begin{figure}[t]
   \centering
 \includegraphics[width=\textwidth,angle=0]{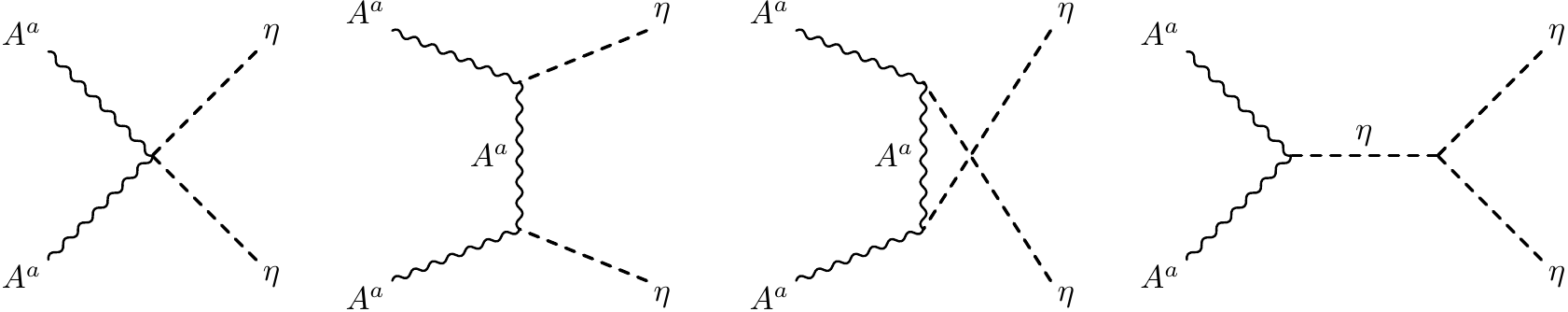}
    \caption{Dark gauge boson annihilation diagrams included in the
    relic density estimate presented in the text.}
    \label{fig:diags}
\end{figure}

Let us now consider the SU(2)$_D$ gauge boson interactions in the model,
\begin{equation}
  \mathcal{L}_{SU(2)_D} =-\frac{1}{4}\left( F^a_{\mu\nu}\right)^2+\left|D_\mu\Phi \right|^2 \,\,\, ,
\label{lsu2}
\end{equation} 
where $F_{\mu\nu}^a=\partial_\mu A^a_\nu -\partial_\nu A^a_\mu + g_D\epsilon^{abc}A_\mu^b A_\nu^c$ and $D_\mu=\partial_\mu-i g_D T^a A_\mu^a$.
The second term of (\ref{lsu2}) contains interactions between $\varphi$ and the $A^a$ gauge fields:
\begin{equation}
\lag_{SU(2)_D} =-\frac{1}{4}\left( F^a_{\mu\nu}\right)^2+\frac{1}{2}\left|\partial_\mu\varphi \right|^2+\frac{1}{8}
g_D^2A^a_\mu A^{a\mu}(v_D+\varphi)^2 \,\,\, .
\label{eq:lsu22}
\end{equation} 
Eq.~(\ref{eq:lsu22}) exhibits a non-anomalous SO(3) symmetry under which the three gauge bosons
transform as a triplet; the other particles in the model are singlets under this symmetry.  As pointed out
in Refs.~\cite{vdm}, this SO(3) symmetry is responsible for preserving the stability of the dark gauge boson multiplet.  
If higher-dimension operators were present, this symmetry could be broken, leading to a decaying dark matter
scenario; this possibility is discussed in the second paper of Ref.~\cite{vdm}. Such dimensionful operators cannot
be introduced here due to the assumption of classical scale invariance.   After re-expressing $\varphi$ and $\sigma_0$ in 
terms of the mass eigenstates $h$ and $\eta$,  one may isolate the leading diagrams that are responsible for dark gauge 
boson annihilation; in the case of small mixing angle $\theta$ (which is the relevant limit, given the results of the 
previous section), one obtains a reasonable approximation by considering the diagrams shown in Fig.~\ref{fig:diags}.  
These diagrams are relevant provided that the second Higgs field $\eta$ remains in thermal equilibrium with the ordinary
standard model particle content up to the point at which dark gauge boson freeze-out occurs.  We will come back to this point later.
For the purposes of our relic density estimate, we omit diagrams that change dark matter number by only one unit, i.e., 
$AA\to A\eta$, the same assumption made in the first paper of Ref.~\cite{vdm}.  For the parameter region in which we obtain 
the desired $\Omega_D h^2$, the Higgs portal coupling $\lambda_P \agt 0.001$; in the second paper of Ref.~\cite{vdm}, it was 
found for similar Higgs portal couplings that the omitted diagrams did not substantially affect the relic density estimate; we leave 
their inclusion, as well as sub-leading diagrams that change dark matter number by two units, for a more detailed analysis in 
future work.

We find that the thermally averaged annihilation cross section times relative velocity that follows from Fig.~\ref{fig:diags} is
\begin{align}
 \langle\sigma_{\text{ann}} v\rangle=&\frac{g_D^4 \cos^4\theta}{192\pi m_A^2}\sqrt{1-\frac{m_\eta^2}{m_A^2}}\left[\left
 (\frac{3}{2}\frac{\lambda \langle\varphi\rangle^2\cos^2\theta }{\left(4m_A^2-m_\eta^2\right)}+\frac{1}{2}\right)^2\right.
 \nonumber\\
 &\left.-\frac{4}{3} \left(\frac{m_A^2}{m_\eta^2-2 m_A^2}\right)^2 \left(8-6\frac{m_\eta^2}{m_A^2}+\frac{m_\eta^4}
 {m_A^4}\right) \left(\frac{3}{2}\frac{\lambda \langle\varphi\rangle^2\cos^2\theta}{\left(4
   m_A^2-m_\eta^2\right)}+\frac{1}{2}\right)\right.\nonumber\\
   &\left.+\frac{4}{3}\left(\frac{m_A^2}{m_\eta^2-2 m_A^2}\right)^2 \left(6-4\frac{m_\eta^2}{m_A^2}
   +\frac{m_\eta^4}{m_A^4}\right)\right].
\end{align}
From this result, the freeze-out temperature and relic density are numerically calculated. With $x \equiv m_A/T$,
we find numerically that the freeze-out temperature is typically in the range $x_F \approx 26 - 27$.   The relic
density is given by
\begin{equation}
\Omega_D h^2 \approx 3 \cdot \frac{(1.07 \times 10^9 \mbox{ GeV}^{-1}) \, x_F}
{\sqrt{g_*(x_F)} M_{Pl} \langle \sigma v \rangle_F}
\end{equation}
where the factor of $3$ takes into account the size of the SU(2)$_D$ gauge multiplet.    As a point of reference,
we note that if all species are dynamical and in equilibrium, one would find $g_*=122$; we take into account
the temperature dependence of $g_*$ in our numerical analysis.

\begin{figure}[t]
    \includegraphics[width=0.495\textwidth]{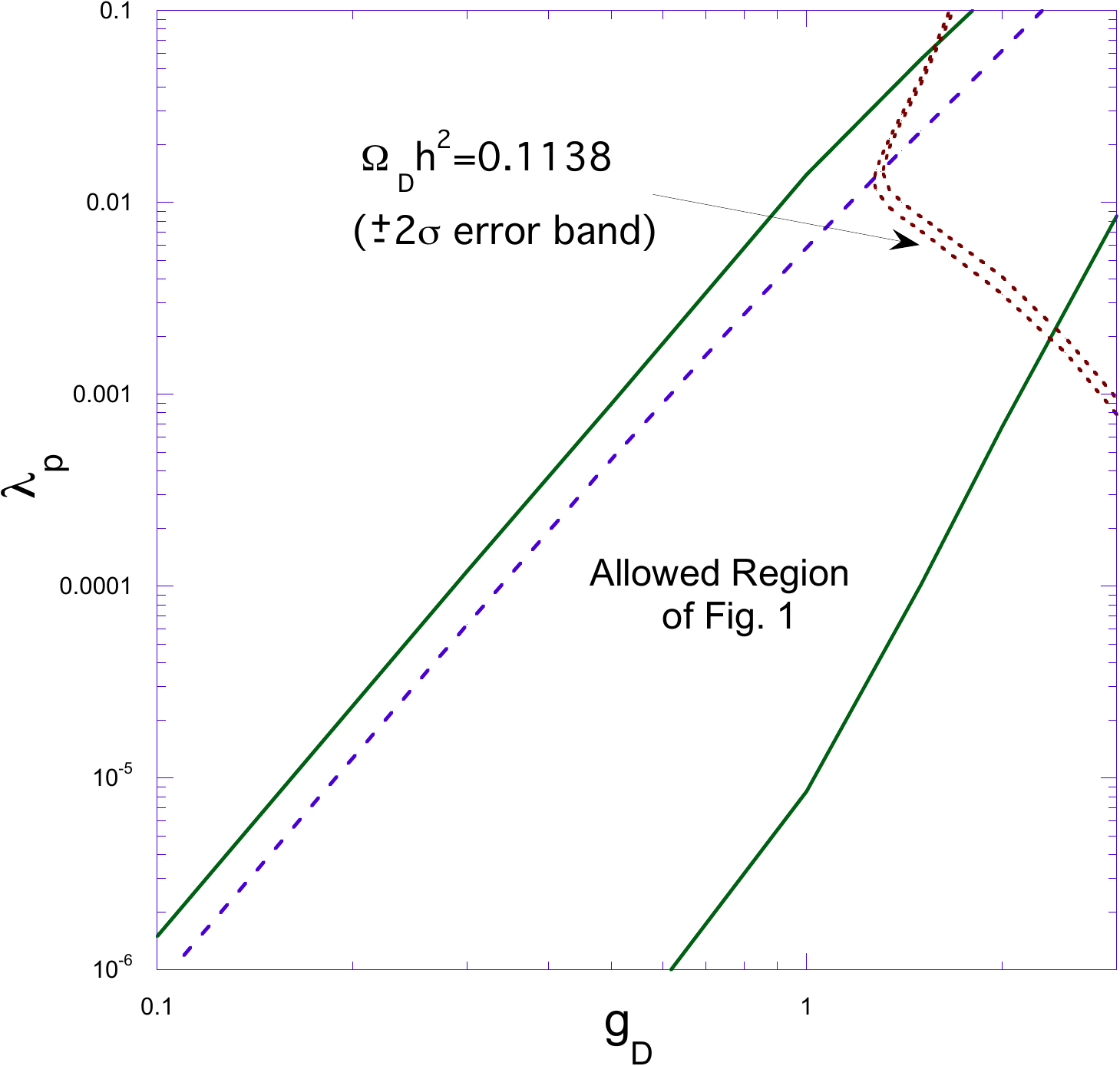} 
    \caption{Band where the dark gauge multiplet provides the dark matter relic density within $\pm2\sigma$ experimental
    uncertainty.}  
    \label{fig:relicallowed}
\end{figure}

The region in parameter space where $0.1048 < \Omega_D h^2 < 0.1228$, the $\pm2\sigma$ band for the WMAP result
$0.1138 \pm 0.0045$~\cite{wmap}, is shown in Fig.~\ref{fig:relicallowed}, together with our previous constraints.
In order to accommodate the observed relic density, the annihilation cross section must be sufficiently
large, which in turn requires larger values of $g_D$ and $\lambda_p$ than allowed if $m_\eta<m_h$.
Hence, our relic density results shown relative to the allowed parameter region of Fig.~\ref{fig:vsp}b.

We note that for all the allowed points in this band, the $\eta$ remains in thermal equilibrium with the standard model
particle bath at the time that the dark matter freezes out.  The relevant constraint (following from decay and inverse decay) 
is $\Gamma_\eta > H(x_F)$, where $\Gamma_\eta$ is the $\eta$ decay width and $H$ is the Hubble 
parameter~\cite{secluded}; we find that this inequality is satisfied by many orders of magnitude for allowed points in the 
$\Omega_D h^2$ band.  Moreover, we find that the two-into-two process  $\eta\eta \rightarrow h h$ is sufficient for 
maintaining $\eta$ equilibrium by itself, for all points in the $\Omega_D h^2$ band that are also within the previously
allowed region. 

Finally, we check the compatibility of our results with current dark matter direct detection bounds. The dark matter-nucleon 
elastic scattering cross section is given by
\begin{equation}
\sigma(NA\to NA)=\frac{1}{64\pi}f^2g_D^4\sin^2 2\theta \frac{m_N^2}{m_A^2}\frac{\langle\varphi\rangle^2}{\langle\sigma\rangle^2}\frac{(m_\eta^2-m_h^2)^2}{m_\eta^4m_h^4}\left(\frac{m_Nm_A}{m_N+m_A}\right)^2
\label{eq:ddsig}
\end{equation}
where $m_N$ is the nucleon mass and $f$ parameterizes the Higgs-nucleon coupling.  In Table~\ref{table1}, we provide more
detailed information on a sampling of points within the  $\Omega_D h^2$~allowed band of Fig.~\ref{fig:relicallowed}, including the
direct detection cross section. The table displays results for $f=0.3$; for different choices of $f$, the results can be scaled 
according to Eq.~(\ref{eq:ddsig}). All the points shown are consistent with the bounds from the Xenon100 
experiment~\cite{xenon100}.  We find the same to be true for all points in the  $\Omega_D h^2$~allowed band 
above $g_D \approx 1.23$.
\begin{table}[t]
\begin{tabular}{cccccccc}
\hline\hline
$g_D$ &  \hspace{0.7em} $\lambda_p$ ($\times 10^{-3}$) \hspace{0.7em}& & \hspace{0.7em}$\langle\varphi\rangle$ 
(GeV)\hspace{0.7em} &\hspace{0.7em} $m_A$ (GeV)\hspace{0.7em} & \hspace{0.7em} $m_\eta$ (GeV) \hspace{0.7em}& \hspace{0.7em}$\sin 
\theta$ \hspace{0.7em}& \hspace{0.7em} $\sigma(AN)$ ($\times 10^{-45}\text{cm}^2$) \hspace{0.7em} \\
\hline
1.4 & 9.127 & &1410 & 987 & 235   & 0.0802 &  1.279 \\
1.5 & 7.689 & &1531 & 1148 & 292 & 0.0417   & 0.5176 \\
2.0 & 3.609 & &2228 & 2228 & 752 & 0.0036   & 0.00972 \\
2.5 & 1.795 & &3158 & 3947& 1666 & 0.0005  & 0.00031 \\
3.0 & 0.8606  & & 4561 & 6841 & 3465 & 0.00008   & 0.00001 \\
\hline\hline
\end{tabular}
\caption{Sample points with $\Omega_Dh^2=0.1138$, the central WMAP value~\cite{wmap} used in Fig.~\ref{fig:relicallowed}.  All
points shown have an elastic scattering cross section $\sigma(AN)$  below the current Xenon100 direct detection 
bounds~\cite{xenon100}.}
\label{table1}
\end{table}

It is now easier to see why this model can be categorized as a secluded dark matter scenario~\cite{secluded}.  The dark 
matter annihilates to an unstable mediator particle, $\eta$, at a rate controlled primarily by the
coupling $g_D$.  On the other hand, the direct detection cross section, Eq.~(\ref{eq:ddsig}), can be made small independently,
by choosing $\lambda_p$ values at fixed $g_D$ that produce small $\sin^2 2\theta$.  Table~\ref{table1} indicates this behavior as 
one moves along the $\Omega_d h^2$ band toward the right side of Fig.~\ref{fig:relicallowed}.

\section{Conclusions} \label{sec:conc}

We have investigated an extension of the standard model that is classically scale-invariant and in which
the electroweak scale arises via the Coleman-Weinberg mechanism~\cite{cw}.   Like similar models involving
new Abelian gauge groups~\cite{recent}, our non-Abelian model communicates the dimensional transmutation that
originates in a dark sector to standard model particles via the Higgs portal.  We
have shown that there are regions of the model parameter space in which the theory maintains
vacuum stability and perturbativity between the electroweak and the Planck scales, and in which
the modifications to the Higgs sector would not yet have been discerned at the LHC.  We have
also shown that the particular gauge extension we discuss provides a dark matter candidate,
a multiplet of stable vector bosons which behaves in accord with secluded dark matter scenarios~\cite{secluded}
that have been discussed in the literature.  

We note that modifications of this model may also be of interest.  For example, if one wanted a similar
non-Abelian scenario with fermionic rather than vector dark matter, then one could introduce dark 
fermions that obtain masses only via spontaneous SU(2)$_D$ breaking (so as not to introduce any new 
fundamental mass scale) and provide a decay channel for the dark gauge boson multiplet.  In such 
a scenario, a new fermion could be a potential dark matter matter candidate.  And as indicated 
earlier, one might entertain weakening the constraints we've considered by taking the gravitational cut off of 
the theory to be lower than the conventional Planck scale.  Many other variations of the model and
the analysis are conceivable.

In light of the current LHC data, the origin of the electroweak scale and the nature of the hierarchy 
problem merit an exploration of the widest range of theoretical possibilities, including the classically 
scale-invariant scenarios that have re-emerged as a possibility in the recent literature~\cite{recent} and motivate the present 
work.  In a few years, the LHC may provide more definitive guidance on whether the one of the more popular theoretical 
proposals or a less expected paradigm is relevant in describing physics at the TeV scale.

\vspace{1em}
{\em Note Added:}  After our manuscript was made public, we learned of  work in another recent preprint that 
also considers an SU(2) vector dark matter model in a scale-invariant context: see Ref.~\cite{hsrecent}.  

\begin{acknowledgments}  
This work was supported by the NSF under Grant PHY-1068008.  In addition, C.D.C. thanks Joseph J. Plumeri II for his generous support.
\end{acknowledgments}


\end{document}